\documentclass[12pt]{article}
\usepackage{a4wide}

\usepackage[intlimits]{amsmath}
\usepackage{amstext,amsfonts,amsbsy,eucal,amssymb}
\usepackage{epsfig}
\usepackage{psfrag}
\usepackage{amsthm}
\usepackage{amscd}

\DeclareMathOperator{\fstr}{Str}
\begin{document}
\vspace*{30pt}
{\LARGE Fermionic ${R}$-operator approach 
for the small-polaron model with open boundary condition
} \\
\vspace{12pt}
\begin{center}
       Yukiko \textsc{Umeno}
        \footnote[2]{E-mail: 
                        \texttt{umeno@monet.phys.s.u-tokyo.ac.jp}
                   },
        Heng \textsc{Fan}
        \footnote[8]{E-mail:
                        \texttt{fan@monet.phys.s.u-tokyo.ac.jp}
                   }   
        and
        Miki \textsc{Wadati}
\\
         Department of Physics, Graduate School of Science,\\
         University of Tokyo,\\
         Hongo 7-3-1, Bunkyo-ku, Tokyo 113-0033 
\end{center}
\vspace{12pt}
\vspace{5mm}
Exact integrability and algebraic Bethe ansatz 
of the small-polaron model 
with the open boundary condition 
are discussed
in the framework of the quantum inverse scattering method (QISM).
We employ a new approach where
the fermionic $R$-operator which consists of fermion operators 
is a key object.
It satisfies the Yang-Baxter equation and 
the reflection equation with its corresponding $K$-operator.
Two kinds of 'super-transposition' for the fermion operators
are defined
and the dual reflection equation is obtained.
These equations prove
the integrability 
and the Bethe ansatz equation 
which agrees with the one obtained from 
the graded Yang-Baxter equation and the graded reflection equations.

\section{Introduction}

There have been reported various one-dimensional (1D) integrable models.
In 1D systems, some physical properties
are beyond the perturbative interpretation \cite{Lee}
and therefore integrable models have attracted much interest.
Among the integrable fermion models,
the simplest but non-trivial one is the small-polaron model,
a spinless fermion model
with a hopping term and an interaction term.
It describes the motion of an additional electron in a polar crystal
\cite{Makhankov}
and exhibits some properties that 
differ from those of the Fermi liquid. 

The integrability and other properties 
for the small-polaron model with open boundary condition
have been studied in many ways.
By use of the coordinate Bethe ansatz and the string hypothesis
the critical value of attractive boundary potential
for a boundary bound state was shown.
\cite{Chen}
In the quantum inverse scattering method (QISM),
the integrability and the Bethe ansatz equation
were derived from
the graded Yang-Baxter equation and 
the graded reflection equations. \cite{Zhou1,Zhou2,Fan}

In the QISM, the graded Yang-Baxter equation plays an important role
for the fermion models.
The corresponding spin models 
through the Jordan Wigner transformation
are often useful.
Using the correspondence, we obtain
the graded Yang-Baxter equation 
from the Yang-Baxter equation of the spin model. \cite{Pu,Wadati}
There, the graded Grassmann product and the super-trace are introduced.
We know that the small-polaron model 
is related with
the $XXZ$ Heisenberg model. 

Recently we introduced a fermionic $R$-operator.
It clarifies the integrable structure
of the fermion models with the periodic boundary condition.
The fermionic $R$-operator satisfies the Yang-Baxter equation.
The 'super-trace' of the fermion operator is introduced
and the commutativity of the transfer operator
proves the integrability. 
\cite{Umeno1,Umeno2}
Furthermore this fermionic $R$-operator enables us to evaluate
thermodynamic quantities for the fermion system 
through the quantum transfer matrix (QTM) method.
In the QTM method,
partition function is estimated in two dimensional space
(the quantum space and the auxiliary space).
The transfer operator exhibits the integrable structure 
and the QTM leads to the partition function.
The former operates on the quantum space,
on the other hand, the latter operates on the auxiliary space. 
Because the fermionic $R$-operator has the same footing on both spaces,
the QTM method can be applied. \cite{Umeno3}

In this paper 
we extend the fermionic $R$-operator approach
to the open boundary condition.
In \S 2, we briefly review the periodic boundary case
of the small-polaron model
with this $R$-operator approach,
where 'super-trace' ($\fstr$) is introduced.
The Yang Baxter equation leads to the integrability and 
the Bethe ansatz equation.
We also investigate the properties of the fermionic $R$-operator
and define two kinds 
of 'super-transposition' (${\rm st}$ and $\bar{\rm st}$).
In \S 3, we prove the integrability 
with the open boundary condition.
We obtain the reflection equation and the dual reflection equation
and prove the commutativity of the transfer operator. 
In \S 4, we evaluate the eigenvalue of the transfer operator
and get the corresponding Bethe ansatz equation.
This result agrees with the one obtained by
the graded Yang Baxter approach.
The last section is devoted to the concluding remarks.

\section{Small-polaron model with periodic boundary condition}

We summarize the fermionic $R$-operator approach 
to the small-polaron model with the periodic boundary condition.
The Hamiltonian of this system is 
\begin{align}
&H = \sum_{j=1}^N H_{j j+1},
\label{hampbc}
\\
&H_{j j+1} = - t \{ c^{\dagger}_{j+1} c_j + c^{\dagger}_j c_{j+1} \}
            + V n_j n_{j+1},
\label{hamdenspbc}
\end{align}
where $t$ represents the overlapping integral and
$V$ 
the electron-phonon coupling.  
The fermionic $R$-operator,
\begin{align}
R_{ab}(u) &= 
a(u) \{ - n_a n_b + \bar{n}_a \bar{n}_b \}] +
b(u) \{ n_a \bar{n}_b + \bar{n}_a n_b \} +
c(u) \{ c^{\dagger}_a c_b + c^{\dagger}_b c_a\},
\label{fer-r}
\\
&a(u) = \dfrac{\sin(u+2\eta)}{\sin 2\eta}, \hspace{2mm} 
b(u) = \dfrac{\sin u}{\sin 2\eta}, \hspace{2mm} 
c(u) =1,
\end{align}
consists of fermion operators
where $n_a=c_a^{\dagger}c_a$ and $\bar{n}_a = 1-n_a$. 
This $R$-operator satisfies the Yang Baxter equation,
\begin{align}
R_{12}(u-v) R_{13}(u) R_{23}(v) 
= R_{23}(v) R_{13}(u) R_{12}(u-v).
\label{yb}
\end{align}
As is the routine for
the quantum inverse scattering method, 
from the equation (\ref{yb})
we have the global Yang Baxter equation,
\begin{align}
&R_{ab}(u-v) T_a(u) T_b(v) 
= T_b(v) T_a(u) R_{ab}(u-v),
\label{globyb}
\end{align}
where the monodromy operator is expressed by 
\begin{align}
T_a(u) &= R_{aN} \ldots R_{a1}
\\
&= A(u) \bar{n}_a + B(u) c_a + C(u) c^{\dagger}_a + D(u) n_a.
\label{monod}
\end{align}
Here the suffix $1, 2 \ldots N$ mean the quantum space
and the suffix $a, b$ 
the auxiliary space.
This equation (\ref{globyb})
leads to the commutativity of the transfer operators $\tau(u)$,
\begin{align}
&[\tau(u), \tau(v)] = 0,
\\ 
&\hspace{5mm}
\tau(u) = \fstr_a T_a(u)
= A(u) - D(u).
\end{align}
Here 'super-trace' for the fermion operators ($\fstr$) is defined as 
\begin{align}
\fstr_a X_a = {}_a\langle 0 | X_a  | 0 \rangle_a
            - {}_a\langle 1 | X_a  | 1 \rangle_a ,
\end{align}
where $|0\rangle_a$ and $|1\rangle_a$ are the fermion Fock states
satisfying $|1\rangle_a = c^{\dagger}_a |0\rangle_a$ and
$|0\rangle_a = c_a |1\rangle_a$. 
This means that the system has a sufficient number of conserved operators
and therefore is integrable.
The logarithmic derivative of the transfer operator at $u=0$
gives the Hamiltonian defined by
(\ref{hampbc}) and (\ref{hamdenspbc})
with 
\begin{align}
t = - \dfrac{1}{\sin 2\eta},
\hspace{3mm}
V = \dfrac{\cos 2\eta}{\sin 2\eta}. 
\label{tV}
\end{align}
The energy spectrum is obtained from the eigenvalue of the transfer operator.
The eigenstate $|\phi\rangle$ and 
the eigenvalue $\Lambda(u)$ of the transfer operator are given by
\begin{align}
&\tau(u)|\phi\rangle=\Lambda(u)|\phi\rangle,
\hspace{3mm}
|\phi\rangle=\prod_{j=1}^M B(u_j) |0\rangle,
\label{eigenstate}
\\
&\Lambda(u)= a(u)^N \prod_{j=1}^M \dfrac{a(u_j-u)}{b(u_j-u)}
             - b(u)^N \prod_{j=1}^M \dfrac{a(u-u_j)}{b(u-u_j)},
\end{align}
where $|0\rangle$ is a pseudovacuum state.
There is a constraint on $u_j (j=1 \ldots M)$ (the Bethe ansatz equation)
so that (\ref{eigenstate}) should be satisfied,
\begin{align}
\left( \dfrac{a(u_i)}{b(u_i)} \right)^N
= \prod_{j \neq i, j=1}^M
  \dfrac{a(u_i-u_j)}{a(u_j-u_i)},
\label{betheansatzpbc}
\end{align}
In terms of the solution $\{ u_j \}$ of (\ref{betheansatzpbc}),
the energy spectrum is given by
\begin{align}
E = \dfrac{\rm d}{{\rm d}u} \log{\Lambda(u)} |_{u=0}
  = N \cot 2\eta 
    - \sum_{j=1}^M\dfrac{\sin 2\eta}{\sin u_j \sin(u_j+2\eta)} .
\label{energypbc}
\end{align}

The fermionic $R$-operator has important properties 
which are useful further in the analysis.
We define two kinds of 'super-transposition'
for the fermion operators ($\rm st$ and $\bar{\rm st}$)
so that 
$(X_a^{{\rm st}_a})^{\bar{\rm st}_a} = X_a$
for any fermion operator $X$,
\begin{align}
&X_a  = A \bar{n}_a + B c_a + C c^{\dagger}_a  + D n_a,
\\
&X_a^{{\rm st}_a} = A \bar{n}_a + B c^{\dagger}_a - C c_a  + D n_a,
\\
&X_a^{\bar{\rm st}_a} = A \bar{n}_a - B c^{\dagger}_a + C c_a  + D n_a.
\end{align}
The unitarity and the crossing unitarity conditions 
are satisfied as follows,

Unitarity
\begin{align}
R_{12}(u)R_{12}(-u) = \rho(u), 
\hspace{5mm}
\rho(u) = 1-\dfrac{\sin^2 u}{\sin^2 2\eta}, 
\label{unitarity}
\end{align}

Crossing unitarity
\begin{align}
R_{12}(u)^{{\rm st}_1} R_{12}(-u-4\eta)^{{\rm st}_2}
= \tilde{\rho}(u),
\hspace{5mm}
\tilde{\rho}(u)
=1 - \dfrac{\sin^2(u+2\eta)}{\sin^2 2\eta}. 
\label{cros-unitarity}
\end{align}
We also have a property,
\begin{align}
R_{12}(u)^{{\rm st}_1} = R_{12}(u)^{\bar{\rm st}_2}.
\end{align}

In the next section,
we shall discuss the fermionic $R$-operator approach
for the open boundary condition.

\section{Exact integrability with open boundary condition}

To treat the open boundary condition, we consider
the Yang-Baxter equation 
and the reflection equations. 
For the fermionic $R$-operator given by (\ref{fer-r}),
we look for the fermionic $K$-operator
which satisfies 
the reflection equation,
\begin{align}
R_{12}(u-v) K_1(u) R_{12}(u+v) K_2(v)
= K_2(v) R_{12}(u+v) K_1(u) R_{12}(u-v).
\label{ref}
\end{align}
We find that the fermionic $K$-operator
is given by
\begin{align}
K_a(u) = - \dfrac{\sin(u-t)}{\sin t} n_a 
+ \dfrac{\sin(u+t)}{\sin t} \bar{n}_a. 
\label{fer-k}
\end{align}
We have a more general form of the fermionic $K$-operator, but
we here restrict 
the $K$-operator to be  
even operator. 
Then the global reflection equation,
\begin{align}
R_{ab}(u-v) \mathcal{T}_a(u) R_{ab}(u+v) \mathcal{T}_b(v)
= \mathcal{T}_b(v) R_{ab}(u+v) \mathcal{T}_a(u) R_{ab}(u-v),
\label{globref}
\end{align}
is satisfied where the double monodromy operator is
\begin{align}
\mathcal{T}_{a}(u) = T_a(u) 
K_a(u) T_a^{-1}(-u). 
\label{doublemonod}
\end{align}

We also show the dual reflection equation,
\begin{align}
\bar{R}_{12}(-u_-) K^+_1(u)^{{\rm st}_1} 
R_{12}(-u_+-4\eta) K^+_2(v)^{\bar{\rm st}_2}
= K^+_2(v)^{\bar{\rm st}_2} R_{12}(-u_+-4\eta)
 K^+_1(u)^{{\rm st}_1} \bar{R}_{12}(-u_-),
\label{dualref}
\end{align}
where 
$\bar{R}_{ab}(u) = R_{ab}(u) ^{{\rm st}_a \bar{\rm st}_b}$
and $u_\pm=u \pm v$.
This dual reflection equation (\ref{dualref}) 
will be used for a proof of
the commutativity of the transfer operator.
The dual $K$-operator is found to be 
\begin{align}
K_a^+(u) = 
\sin(u+2\eta-t^+)
\bar{n}_a +
\sin(u+2\eta+t^+)
n_a .
\end{align}

Next we prove 
the commutativity of the transfer operators
with different spectral parameters.
We define the transfer operator as
\begin{align}
t(u) = \fstr_a  \{ K^+_a(u) \mathcal{T}_a(u) \}.
\end{align}
We prepare some properties of 
the 'super-trace' and the 'super-transposition' for
the fermion operators. 
When $Y$ is an even fermion operator
and $X$ operates trivially except on $a$ space, we have 
\begin{align}
\{X_a Y_{ab} \}^{{\rm st}_a} 
= Y_{ab}^{{\rm st}_a} X_a^{{\rm st}_a}.
\label{prosuptrans1}
\end{align}
When $X$ and $Y$ are even fermion operators, we have
\begin{align}
\fstr_{ab} \{X_{ab} Y_{ab}\}
= 
\fstr_{ab} \{X_{ab}^{{\rm st}_a} Y_{ab}^{{\rm st}_a}\}.
\label{prosuptrace1}
\end{align}
When $X$ and $Y$ are even fermion operators
and $X$ operates trivially except on $a$ and $b$ spaces, we have 
\begin{align}
\{ X_{ab} Y_{ab} \}^{{\rm st}_a\bar{\rm st}_b}
= Y_{ab}^{{\rm st}_a\bar{\rm st}_b}
X_{ab}^{{\rm st}_a\bar{\rm st}_b},
\label{prosuptrans2}
\\
\fstr_{ab} \{X_{ab} Y_{ab}\}
=\fstr_{ab} \{ Y_{ab} X_{ab}\}.
\label{prosuptrace2}
\end{align}

The relation $t(u)t(v) = t(v)t(u)$
is to be proved. We begin with
\begin{align}
&t(u) t(v)
\\
&= \fstr_a \{ K^+_a(u) \mathcal{T}_a(u) \} 
\fstr_b \{ K^+_b(v) \mathcal{T}_b(v) \}
\\
&= \fstr_a \{ K^+_a(u)^{{\rm st}_a} \mathcal{T}_a(u)^{{\rm st}_a} \} 
\fstr_b \{ K^+_b(v) \mathcal{T}_b(v) \}
\\
&= \fstr_{ab} [ K^+_a(u)^{{\rm st}_a}  \mathcal{T}_a(u)^{{\rm st}_a} 
 K^+_b(v) \mathcal{T}_b(v) ]
\end{align}
Because $K$-operator is even,
$\mathcal{T}_a(u)^{{\rm st}_a}$ commutes with $K^+_b(v)$.
From the crossing unitarity (\ref{cros-unitarity}),
the identity operator can be inserted as 
\begin{align}
&= \tilde{\rho_+}^{-1}
\fstr_{ab} [ K^+_a(u)^{{\rm st}_a}  K^+_b(v) 
R_{ab}(-u_+ - 4\eta)^{{\rm st}_b} 
R_{ab}(u_+)^{{\rm st}_a} 
\mathcal{T}_a(u)^{{\rm st}_a} \mathcal{T}_b(v) ]
\\
&= \tilde{\rho_+}^{-1}
\fstr_{ab} [
\{ K^+_a(u)^{{\rm st}_a} R_{ab}(-u_+ - 4\eta)
 K^+_b(v)^{\bar{\rm st}_b} \}^{{\rm st}_b}
\{\mathcal{T}_a(u) R_{ab}(u_+)
 \mathcal{T}_b(v) \}^{{\rm st}_a} ]
\\
&= \tilde{\rho_+}^{-1}
\fstr_{ab} [
\{ K^+_a(u)^{{\rm st}_a} R_{ab}(-u_+ - 4\eta)
 K^+_b(v)^{\bar{\rm st}_b} \}^{{\rm st}_b \bar{\rm st}_a} 
\{\mathcal{T}_a(u) R_{ab}(u_+)
 \mathcal{T}_b(v) \} ]
\label{conpare1}
\end{align}
The above equalities are due to
the properties (\ref{prosuptrans1}) and (\ref{prosuptrace1})
respectively,
where $u_{\pm}= u \pm v, \rho_{\pm} = \rho(u_{\pm}),
\tilde{\rho}_{\pm} = \tilde{\rho}(u_{\pm})$.
Then from the unitarity (\ref{unitarity})
and property (\ref{prosuptrans2}), we get
\begin{align}
= \tilde{\rho_+}^{-1} \rho_-^{-1}
\fstr_{ab} &[
\{ K^+_a(u)^{{\rm st}_a} R_{ab}(-u_+ - 4\eta)
 K^+_b(v)^{\bar{\rm st}_b} \}^{{\rm st}_b \bar{\rm st}_a} 
\nonumber\\
&R_{ab}(-u_-) R_{ab}(u_-)
\{\mathcal{T}_a(u) R_{ab}(u_+)
 \mathcal{T}_b(v) \} ]
\\
= \tilde{\rho_+}^{-1} \rho_-^{-1}
\fstr_{ab} &[
\{R_{ab}(-u_-)^{{\rm st}_a \bar{\rm st}_b}
 K^+_a(u)^{{\rm st}_a} R_{ab}(-u_+ - 4\eta)
 K^+_b(v)^{\bar{\rm st}_b} \}^{{\rm st}_b \bar{\rm st}_a} 
\nonumber\\
&\{ R_{ab}(u_-) \mathcal{T}_a(u) R_{ab}(u_+) \mathcal{T}_b(v) \}]
\end{align}
Here the reflection equation (\ref{ref}) and
the dual reflection equation (\ref{dualref}) 
can be applied,
\begin{align}
= \tilde{\rho_+}^{-1} \rho_-^{-1}
\fstr_{ab} &[
\{ K^+_b(v)^{\bar{\rm st}_b} 
R_{ab}(-u_+ - 4\eta)
 K^+_a(u)^{{\rm st}_a} 
R_{ab}(-u_-)^{{\rm st}_a \bar{\rm st}_b}
\}^{{\rm st}_b \bar{\rm st}_a} 
\nonumber\\
&\{ \mathcal{T}_b(v) R_{ab}(u_+)\mathcal{T}_a(u) R_{ab}(u_-) \} ]
\\
= \tilde{\rho_+}^{-1} \rho_-^{-1}
\fstr_{ab} &[
R_{ab}(-u_-)
\{
 K^+_b(v)^{\bar{\rm st}_b} 
R_{ab}(-u_+ - 4\eta)
 K^+_a(u)^{{\rm st}_a} 
\}^{{\rm st}_b \bar{\rm st}_a} 
\nonumber\\
&\{ \mathcal{T}_b(v) R_{ab}(u_+)\mathcal{T}_a(u)\} R_{ab}(u_-) ]
\end{align}
For this expression, we do the other way around. 
The property (\ref{prosuptrans2}) is used in the above, and
the property (\ref{prosuptrace2}) and then
the use of the unitarity (\ref{unitarity}) give
\begin{align}
&= \tilde{\rho_+}^{-1} 
\fstr_{ab} [
\{
 K^+_b(v)^{\bar{\rm st}_b} 
R_{ab}(-u_+ - 4\eta)
 K^+_a(u)^{{\rm st}_a} 
\}^{{\rm st}_b \bar{\rm st}_a} 
\{ \mathcal{T}_b(v) 
R_{ab}(u_+)\mathcal{T}_a(u)\} ]
\label{conpare2}
\\
&=t(v)t(u).
\end{align}
Thus,
the commutativity of transfer operator is proved.

We relate the transfer operator $t(u)$
to the Hamiltonian of this system. 
The derivative of the transfer operator 
is considered.
We have $K_a(0)=1$,  $\mathcal{T}_a(0)=1$
and 
\begin{align}
t(0)=\fstr_a K_a^+(0) = -2 \cos 2\eta \sin t^+.
\end{align}
Using these relations, we obtain
\begin{align}
&
t^{\prime}(0)
=\fstr_a K_a^{+ \prime}(0) +
\fstr\{ K_a^+(0) R_{aN}^{\prime}(0)P_{aN} \} +
\fstr K_a^+(0) \sum_{j=1}^{N-1}
R_{j+1 j}^{\prime}(0) P_{j+1 j}
\nonumber\\
& + \fstr_a K_a^+(0) 
K_1^{\prime}(0)+
\fstr_a K_a^+(0) \sum_{j=1}^{N-1} 
P_{j+1 j} R_{j+1 j}^{\prime}(0) +
\fstr_a \{K_a^+(0) P_{aN} R_{aN}^{\prime}(0)\}
\end{align}
Here ${}^\prime$ means the derivative with respect to $u$.
The Hamiltonian density $H_{j j+1}$,
(\ref{hamdenspbc}) with (\ref{tV}),
is expressed as
$H_{j j+1}=R_{j+1 j}^{\prime}(0) P_{j+1 j}=
P_{j+1 j} R_{j+1 j}^{\prime}(0)$.
Thus the Hamiltonian is given by
\begin{align}
H &= \frac{1}{2} t(0)^{-1} 
(t^{\prime}(0)-\fstr_a K_a^{+ \prime}(0))
\label{hamirelobc}
\\
& = \sum_{j=1}^{N-1} H_{j j+1} -
\frac{1}{2} \cot t 
(2n_1-1) +
\frac{1}{2} \cot t^+ 
(2n_N-1)
\end{align}

\section{Algebraic Bethe ansatz}

In this section, 
the eigenvalue and the eigenstate 
of the transfer operator are discussed.
The double monodromy operator (\ref{doublemonod}) is expressed by
\begin{align}
\mathcal{T}_a(u) = \mathcal{A}(u) \bar{n}_a 
+ \mathcal{B}(u) c_a + \mathcal{C}(u) c_a^{\dagger} + \mathcal{D}(u) n_a. 
\label{doublemonod2}
\end{align}
Because the double-monodromy operator is an even operator,
$\mathcal{B}(u)$ should have the property of 
the creation operator.
We construct the eigenvector algebraically.
We introduce the eigenvector of the transfer operator by
\begin{align}
| \Phi \rangle = \prod _{j=1}^{M} \mathcal{B}(u_j) |0 \rangle. 
\end{align}
The double-monodromy operator (\ref{doublemonod2}) is also expressed 
with the monodromy operators (\ref{monod}),
\begin{align}
\mathcal{T}_a(u) &= T_a(u) K_a(u) T_a^{-1}(-u), 
\\
&T_a(u) = A(u) \bar{n}_a + B(u) c_a + C(u) c^{\dagger}_a + D(u) n_a,
\\
&T_a^{-1}(u)
=  \bar{A}(u) \bar{n}_a + \bar{B}(u) c_a 
+ \bar{C}(u) c^{\dagger}_a + \bar{D}(u) n_a.
\end{align}
The above relations give 
$\mathcal{A}(u) \ldots \mathcal{D}(u)$
in terms of $A(u) \ldots D(u)$ and $\bar{A}(-u) \ldots \bar{D}(-u)$.
We have 
the relations between $A(u) \ldots D(u)$
and $\bar{A}(-u) \ldots \bar{D}(-u)$
obtained from the global Yang Baxter equation (\ref{globyb})
with $v=-u$
\begin{align}
T_b^{-1}(-u) R_{ab}(2u) T_a(u)
= T_a(u) R_{ab}(2u) T_b^{-1}(-u).
\end{align}
It is 
convenient to use $\tilde{\mathcal{D}}(u)$,
\begin{align}
 \tilde{\mathcal{D}}(u) = \sin(2u+2\eta) \mathcal{D}(u) 
- \sin 2\eta \mathcal{A}(u)
\end{align}
instead of $\mathcal{D}(u)$.
The eigenvalue of $\tilde{\mathcal{D}}(u)$ 
acting on the pseudovacuum $|0\rangle$ takes a
simpler form than that of $\mathcal{D}(u)$,
\begin{align}
&\mathcal{A}(u) |0 \rangle = \dfrac{\sin(u+t)}{\sin t} a(u)^{2N} |0 \rangle,
\\
&\tilde{\mathcal{D}}(u) |0 \rangle = 
- \dfrac{\sin 2u \sin(u+2\eta-t)}{\sin t} b(u)^{2N} |0 \rangle.
\end{align}
The transfer operator is rewritten as
\begin{align}
t(u) &= \fstr K_a^+(u) \mathcal{T}_a(u)
\\
     &= \sin(u+2\eta+t^+) \mathcal{A}(u) - \sin(u+2\eta-t^+)
     \mathcal{D}(u)
\\
&= - \dfrac{\sin(2u+4\eta) \sin(u-t^+)}{\sin(2u+2\eta)} \mathcal{A}(u)
+ \dfrac{\sin(u+2\eta+t^+)}{\sin(2u+2\eta)} \tilde{\mathcal{D}}(u).
\end{align}

The global reflection equation 
gives the relations among $\mathcal{A}(u) \ldots \mathcal{D}(u)$.
For instance, we have
\begin{align}
&\mathcal{B}(u)\mathcal{B}(v)=\mathcal{B}(v)\mathcal{B}(u),
\\
&a_1 b_2 \mathcal{B}(u)\mathcal{A}(v) 
= b_1 c_2 \mathcal{C}(v)\mathcal{D}(u) + c_1 b_2 \mathcal{C}(v)\mathcal{D}(u)
     + b_1 a_2 \mathcal{A}(v)\mathcal{B}(u), 
\\
&c_1 a_2 \mathcal{B}(u)\mathcal{D}(v) 
+ b_1 b_2 \mathcal{D}(u)\mathcal{B}(v) + c_1 c_2 \mathcal{A}(u)\mathcal{B}(v)
= a_1 a_2 \mathcal{B}(v)\mathcal{D}(u) + a_1 c_2 \mathcal{A}(v)\mathcal{B}(u),
\\
&a_1 c_2 \mathcal{B}(u)\mathcal{D}(v) + a_1 a_2 \mathcal{A}(u)\mathcal{B}(v)
= c_1 c_2 \mathcal{B}(v)\mathcal{D}(u) + b_1 b_2 \mathcal{B}(v)\mathcal{A}(u) 
+ c_1 a_2 \mathcal{A}(v)\mathcal{B}(u),
\end{align}
where $a_1= a(u-v)$, $a_2=a(u+v), \ldots$ and so on.
Using $\tilde{\mathcal{D}}(u)$
instead of $D(u)$, we obtain
\begin{align}
\tilde{\mathcal{D}}(u)\mathcal{B}(v)
&=\dfrac{\sin(u+v+4\eta)\sin(u-v+2\eta)}
{\sin(u-v)\sin(u+v+2\eta)} \mathcal{B}(v)\tilde{\mathcal{D}}(u)
- \dfrac{\sin 2\eta \sin 2(u+2\eta)}
{\sin(u-v)\sin 2(v+\eta)} \mathcal{B}(u)\tilde{\mathcal{D}}(v)
\nonumber\\
&+ \dfrac{\sin 2\eta \sin 2v \sin 2(u+2\eta)}
{\sin(u+v+2\eta)\sin 2(v+\eta)} \mathcal{B}(u)\mathcal{A}(v),
\\
\mathcal{A}(u)\mathcal{B}(v)
&=\dfrac{\sin(u-v-2\eta)\sin(u+v)}{\sin(u-v)\sin(u+v+2\eta)} 
\mathcal{B}(v)\mathcal{A}(u)
-\dfrac{\sin 2\eta}{\sin(u+v+2\eta)\sin 2(v+\eta)} 
\mathcal{B}(u)\tilde{\mathcal{D}}(v)
\nonumber\\
&+\dfrac{\sin 2\eta \sin 2v}{\sin(u-v)\sin 2(v+\eta)} 
\mathcal{B}(u)\mathcal{A}(v).
\end{align}
These relations lead to the eigenvalue
of the transfer operator, 
\begin{align}
\Lambda (u) 
&=- \dfrac{\sin(2u+4\eta) \sin(u-t^+)}{\sin(2u+2\eta)}
    \dfrac{\sin(u+t)}{\sin t}
   a(u)^{2N} 
  \prod_{j=1}^M 
  \dfrac{\sin(u-u_j-2\eta) \sin(u+u_j)}
        {\sin(u-u_j) \sin(u+u_j+2\eta)}
\nonumber\\
& - \dfrac{\sin(u+2\eta+t^+)}{\sin(2u+2\eta)}
    \dfrac{\sin 2u \sin(u+2\eta-t)}{\sin t}
   b(u)^{2N} 
  \prod_{j=1}^M 
  \dfrac{\sin(u+u_j+4\eta) \sin(u-u_j+2\eta)}
        {\sin(u-u_j) \sin(u+u_j+2\eta)},
\end{align}
and the Bethe ansatz equations ($l=1, \dots M$),
\begin{align}
\left\{ \dfrac{a(u_l)}{b(u_l)} \right\}^{2N}
= \dfrac{\sin(u_l+2\eta+t^+) \sin(u_l+2\eta-t)}
        {\sin(u_l-t^+) \sin(u_l+t)}
  \prod_{j \neq l, j=1}^M
  \dfrac{\sin(u_l+u_j+4\eta) \sin(u_l-u_j+2\eta)}
        {\sin(u_l-u_j-2\eta) \sin(u_l+u_j)}.
\label{betheansatzobc}
\end{align}
From the relation (\ref{hamirelobc}), the energy spectrum is 
\begin{align}
E= - (N-1) \cot 2\eta - \dfrac{1}{2} \cot t + \dfrac{1}{2} \cot t^+
+\dfrac{1}{\sin 4\eta}
- 2 \sum_{j=1}^M \dfrac{\sin 2\eta}{\sin u_j \sin(u_j+2\eta)}
\label{enegyobc}
\end{align}
\section{Concluding Remarks}

In this paper, we have extended the fermionic $R$-operator approach
for the small polaron model with the open boundary condition.
The Yang-Baxter equation and the reflection equations 
are satisfied among 
the fermionic $R$-operators and the fermionic $K$-operators
which consist of fermion operators.
The properties of the fermionic $R$-operators and
the 'super-trace' and the 'super-transposition'
for the fermion operators 
play an important role.
This approach 
gives a simpler and clearer description
than the usual graded Yang-Baxter approach.
The integrability and the Bethe ansatz equation are obtained
in a unified manner.

We discuss future problems related to this work.
First, we solve the Bethe ansatz equations numerically 
to get the physical quantities.
Second, we further extend the fermionic $R$-operator approach 
to the boundary impurity problems.
Last but not least,
we develop the quantum transfer matrix method
for open boundary condition.
We note that this problem remains to be open.

\vspace{5mm}
\begin{center}
{\bf Acknowledgment}
\end{center}
The authors are grateful to M. Shiroishi
for useful discussions and valuable comments.
Y. U. and H. F. acknowledge JSPS Fellowships.

\vspace{30pt}
\begin{flushleft}
APPENDIX A: Comparison with the spin system
\end{flushleft}
\setcounter{equation}{0}
\renewcommand{\theequation}{A.\arabic{equation}}

The small polaron model corresponds to
the $XXZ$ Heisenberg model through the Jordan Wigner transformation.
For the periodic boundary condition (p.b.c.), 
the fermion system and the spin system 
are not exactly the same.
The energy spectrum and its Bethe ansatz equation
for the fermion system are given by 
(\ref{betheansatzpbc}) and (\ref{energypbc})
while those for the spin system are given by 
\begin{align}
&E = \dfrac{\rm d}{{\rm d}u} \log{\Lambda(u)} |_{u=0}
  = N \cot 2\eta 
    - \sum_{j=1}^M\dfrac{\sin 2\eta}{\sin u_j \sin(u_j+2\eta)}, 
\\
&\left( \dfrac{a(u_i)}{b(u_i)} \right)^N
= \prod_{j \neq i, j=1}^M 
\left( - \dfrac{a(u_i-u_j)}{a(u_j-u_i)} \right).
\label{betheansatzpbcspin}
\end{align}
We see that the right hand side of (\ref{betheansatzpbc}) and 
(\ref{betheansatzpbcspin}) are different in signs.
It reflects the difference 
between p.b.c. of fermion system and that of spin system
caused by non-locality of the Jordan Wigner transformation.

On the other hand,
the small polaron model with open boundary condition
is equivalent to 
the XXZ Heisenberg model with boundary magnetic field
and their energy spectrums are exactly the same. 


\end{document}